\documentclass[12]{iopart}

\usepackage{graphicx}
\usepackage{hyperref}
\usepackage[english]{babel}
\usepackage{makeidx}

\begin{document}

\title{Cation composition effects on oxide conductivity in the Zr$_2$Y$_2$O$_7$-Y$_3$NbO$_7$ system.}

\author {Dario Marrocchelli$^1$, Paul A Madden$^2$, Stefan T Norberg$^{3,4}$ and Stephen Hull$^{3}$}

\address{$^1$ School of Chemistry, University of Edinburgh, Edinburgh EH9 3JJ, United Kingdom}
\address{$^2$ Department of Materials, University of Oxford, Parks Road, Oxford OX1 3PH, United Kingdom}
\address{$^3$ The ISIS Facility, Rutherford Appleton Laboratory, Chilton, Didcot, Oxfordshire, OX11 0QX, United Kingdom}
\address{$^4$ Department of Chemical and Biological Engineering, Chalmers University of Technology, SE-412 96 Gothenburg, Sweden}

\ead{\mailto {D.Marrocchelli@sms.ed.ac.uk}}

\begin{abstract}
\newline
{\small  Realistic, first-principles-based interatomic potentials have been used in
molecular dynamics simulations to study the effect of cation composition on the
ionic conductivity in the Zr$_2$Y$_2$O$_7$-Y$_3$NbO$_7$ system and to link the
dynamical properties to the degree of lattice disorder. Across the composition
range, this system retains a disordered fluorite crystal structure and the vacancy
concentration is constant. The observed trends of decreasing conductivity and
increasing disorder with increasing Nb$^{5+}$ content were reproduced in
simulations with the cations randomly assigned to positions on the cation
sublattice. The trends were traced to the influences of the cation charges and
relative sizes and their effect on vacancy ordering by carrying out
additional calculations in which, for example, the charges of the cations were
equalised. The simulations did not, however, reproduce all the observed properties,
particularly for Y$_3$NbO$_7$. Its conductivity was significantly overestimated and
prominent diffuse scattering features observed in small area electron diffraction
studies were not always reproduced. Consideration of these deficiencies led to a
preliminary attempt to characterise the consequence of partially ordering the
cations on their lattice, which significantly affects the propensity for vacancy
ordering. The extent and consequences of cation ordering seem to be much less
pronounced on the Zr$_2$Y$_2$O$_7$ side of the composition range.}
\end{abstract}

\submitto{\JPCM}

\section{Introduction}

Oxide-ion conducting  materials are the subject of intensive research activity due
to their technological applications in Solid Oxide Fuel Cells (SOFCs), oxygen
separation membranes and gas sensors. Cubic, fluorite-structured, binary compounds
of stoichiometry $A$O$_2$  are of particular interest, especially when some of the
host cations are replaced by species of a lower valence to produce anion-deficient
phases like $A^{4+}_{1-x}$$B^{3+}_{x}$O$_{2-x/2}$. These compounds have charge
compensating vacancies which allow the remaining anions to move more rapidly
throughout the crystal and this leads to impressive values of the ionic
conductivity ($\sigma > 0.1\; \Omega^{-1} \rm{cm^{-1}}$, at $T=1500$ K) as observed
in $M_2$O$_3$-doped ZrO$_2$ ($M$ = Sc or Y) and $M_2$O$_3$-doped CeO$_2$ ($M$ = Y or Gd).
\newline

At the simplest level the ionic conductivity would be expected to increase with the
number of vacancies, but this is not the case. In
Zr$^{4+}_{1-x}$Y$^{3+}_{x}$O$_{2-x/2}$ the conductivity decreases with increasing
$x$ for $x>0.16$ and, furthermore, for a given $x$ the value of the conductivity
depends on the identity of the dopant cation \cite{SHu1}. For example,
Zr$^{4+}_{1-x}$Sc$^{3+}_{x}$O$_{2-x/2}$ is a better conductor than
Zr$^{4+}_{1-x}$Y$^{3+}_{x}$O$_{2-x/2}$ \cite{Etsell}. These effects are due to a
reduced mobility associated with the ordering of the vacancies, which may be
influenced by vacancy-vacancy, vacancy-cation and cation-cation interactions.
Bogicevic and co-workers \cite{Bog1} examined these effects in stabilized zirconias
by using first-principles, zero temperature electronic structure calculations. They
found that vacancy ordering is most strongly governed by vacancy-vacancy
interactions, followed by vacancy-dopant and, weakest of all, dopant-dopant
interactions. The influence of the identity of the dopant could be understood in
terms of a balance between competing electrostatic and elastic effects, which are
associated with differences in the cation charges and radii. They showed that the
best way to minimize vacancy-dopant association is to choose dopants where the
strain term ({\it i.e.} the dopant ionic radius relative to that of the host
cation) counteracts the electrostatic interaction, rather than simply matching the
ionic radii. Some of these factors have also been identified in other studies
\cite{Zac,Kil1,Kri,Dev}. \newline

We have carried out Molecular Dynamics (MD) simulation studies of a similar class of
materials with realistic first-principles based interaction models \cite{Norb}, in
order to examine the influence of these effects at the experimentally relevant
temperatures ($\geq$ 1000 K). The interaction potentials are capable of accurately reproducing 
experimental conductivities {\it etc.}, and long range effects of strain
are incorporated by our use of considerably larger simulation cells (typically 4 x
4 x 4 unit cells but up to 10 x 10 x 10)
 than are accessible in direct first-principles calculations. In the present work we focus
on the ternary system Zr$_{0.5-0.5x}$Y$_{0.5+0.25x}$Nb$_{0.25x}$O$_{1.75}$ which
retains a disordered fluorite structure with a constant vacancy count  as $x$ is
varied (one in 8 oxide ions is ``missing"), which allows us to selectively study
the effect of varying dopant cation charge and ionic radius (Nb$^{5+}$ is
significantly smaller than Y$^{3+}$ since $R_{cat}$ = $0.74$, $0.84$, $1.019$ \AA\
for Nb$^{5+}$ , Zr$^{4+}$ and Y$^{3+}$ respectively \cite{Shannon}). Although the
number of vacancies is fixed across the whole series, the conductivity changes by
almost two orders of magnitude from {\it x} = 0 to {\it x} = 1 at 1000 K \cite{Lee1}.
Furthermore, it was observed that the intensity of diffuse scattering, related to
the lattice disorder, {\em decreases} as the conductivity {\em increases} which
presented the prospect of linking structural (crystallographic) studies to the
effects on the conduction mechanism and the conductivity.

Consequently, we have examined \cite{Norb, Mar} this system with a combination of
impedance spectroscopy, powder neutron diffraction - including the analysis of the
total scattering using bond valence sum constrained Reverse Monte Carlo modelling
(RMC) \cite{McG2, Tuck, NorbergRMC} - and MD simulations. The RMC analysis
indicated that real-space configurations based on a fluorite structure with the
dopant cations randomly arranged within the cation sublattice, {\it i.e.} a
disordered-fluorite ({\it d}-fluorite) structure, gave an excellent representation
of the diffraction data across the composition range. The MD simulations were
initiated from such {\it d}-fluorite configurations. They {\em accurately}
reproduced the average structure extracted from the diffraction data using RMC at
the level of the partial radial distribution and bond-angle distribution functions.
Furthermore, the simulations reproduced the temperature-dependence of the
experimental conductivity for Zr$_2$Y$_2$O$_7$ ({\it i.e.} the $x$=0 system) and
the trend of decreasing conductivity with increasing Nb content across the series.
A substantial part of the present paper will be concerned with a detailed analysis
of these simulations and examination of the chemical factors which contribute to
this trend. These analyses enable us to track the competition between Coulomb and
strain effects on the vacancy ordering in the {\it d}-fluorite structure and to
explain the consequences for the ionic mobility.
\newline

Other considerations suggest, however, that these simulations may miss an important
contributor to the conductivity of these materials. Small area electron diffraction
(SAED) studies  have been carried out on these (and related) materials by several
groups \cite{Miida,Irvine2000,Whittle}. This technique gives single-crystal diffraction information. The SAED
studies show sharp diffuse scattering features at particular positions in
reciprocal space which are the result of a structural modulation which is not
intrinsic to the {\it d-}fluorite structure \cite{Miida,Irvine2000,Whittle}. The
diffuse peaks are associated with a pattern of strain caused by vacancy ordering
and, according to some authors \cite{Miida}, to partial ordering of the dopant
cations. The ordering is of intermediate range, one analysis suggesting a
correlation length of $\sim$22 \AA\ in Y$_3$NbO$_7$ \cite{Miida}. The diffuse peaks
suggest a structure with a local ordering tendency in which the anion vacancies are
aligned in pairs along the $\langle 111\rangle$ directions within the $x=$1 (Y$_3$NbO$_7$)
composition, in a manner related to the pyrochlore structure. As $x\to 0$
(Zr$_2$Y$_2$O$_7$) the diffuse scattering changes and, as interpreted by Irvine
{\it et al} \cite{Irvine2000}, is associated with vacancy alignment along $\langle110\rangle$
directions, which resembles the situation within the C-type structure of
Y$_2$O$_3$. According to these authors, the decrease in conductivity as $x\to$1 is
caused by the effects of these different patterns of vacancy ordering on the anion
mobility, though no clear explanation of this structure-property relationship has
been provided to date. To anticipate the story below somewhat, not all of these peaks are
seen in diffraction patterns calculated from our {\it d-}fluorite simulations,
even if they are initiated with configurations in which the oxygen ions are placed
in positions consistent with the pyrochlore vacancy ordering. We have,
therefore been led to perform a new set of simulations in which we explore the
effect of some degree of local cation ordering. \newline

In this paper, we begin with a detailed examination of the structure and ion dynamics in the {\it
d-}fluorite simulations which successfully reproduce the trends observed in these properties
in the Zr$_{0.5-0.5x}$Y$_{0.5+0.25x}$Nb$_{0.25x}$O$_{1.75}$ series. In order to
clarify how they are affected by the cation properties, we consider the effects of
modifying our first-principles interaction potentials by equalising the cation
charges and the range of the short-range repulsive interactions between the cations
and the oxide ions. We then consider the structure of the material from the
perspective of the vacancies and demonstrate similar cation-specific
vacancy-ordering tendencies to those which have been seen in the ab initio studies
of stabilised zirconias \cite{Bog1}. For the reasons described above, we are then led to
examine the consequences of partially ordering the cations for the energetics, the
diffuse scattering and the conductivity. The studies here are less definitive than
the {\it d-}fluorite ones, as we cannot determine the degree of ordering within our
MD simulations and must rely on drawing conclusions from different postulates for
the order. The partially ordered structures necessarily have a lower entropy than
the fully disordered ones, and therefore become unstable with respect to them at
high temperatures. However it is only at very high temperatures (where these phases
are unstable) that the cations move sufficiently to allow the structure to
equilibrate.

\section{Disorder and mobility  in the {\it d-}fluorite structure of Zr$_2$Y$_2$O$_7$  and Y$_3$NbO$_7$ }
\label{differences}

The simulation methodology is summarised in \ref{Sim-det}. The
simulations are thoroughly equilibrated at high temperatures (1500-2000 K) in
constant pressure simulations before obtaining dynamical information from constant
volume runs at the zero-pressure volume. At these high temperatures the oxide ions
are diffusing on the simulation timescale but there is no significant exchange of
cations. We also quenched down to room temperature and conducted further short runs
in order to examine structural information and compare with diffraction data collected at the
same temperature. Note that, because the ionic mobility drops
sharply as the temperature is reduced during the quench, that these runs are not
fully equilibrated. To some extent, this will also be true of experimental samples
which are prepared at high temperature and quenched, though the rate of cooling is
much slower than for the simulated systems. \newline

We begin by discussing results on the {\it d-}fluorite structure, which are
initialised by randomly distributing the cation species over the cation sublattice and
randomly assigning oxide anions to the anion lattice. Theoretical calculations show
the energy scale associated with cation ordering is very small \cite{Bog1,Sick} in
yttria-stabilised zirconia and, as we showed in the previous paper \cite{Norb}, 
extensive powder diffraction and RMC analyses across the whole series
Zr$_{0.5-0.5x}$Y$_{0.5+0.25x}$Nb$_{0.25x}$O$_{1.75}$ provided no evidence for
cation ordering, in agreement with previous studies \cite{Lee1}. We also
demonstrated the excellent agreement between the  present simulations and the RMC
data, as well as a very good reproduction of the temperature dependence of the
experimental conductivity for Zr$_2$Y$_2$O$_7$. In the present paper we will only
discuss results for the end members ($x$=0 and $x$=1) of the series, {\it i.e.}
Zr$_2$Y$_2$O$_7$ and Y$_3$NbO$_7$. \newline

\subsection{Cation-induced disorder}

\begin{figure}[htbp]
\begin{center}
\includegraphics[width=8cm]{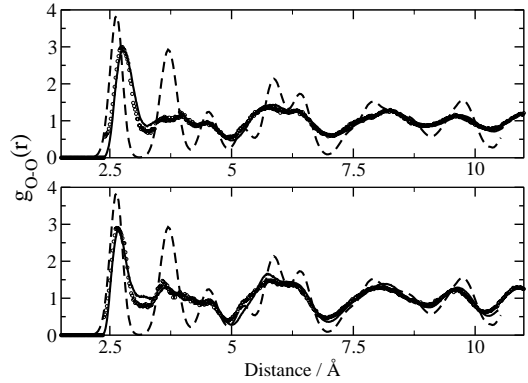}
\end{center}
\caption{\it{\small Comparison between the simulated (solid line)  $g_{\rm{O-O}}(r)$ (obtained from the QUAIM potential
described in \cite{Norb}) in Zr$_2$Y$_2$O$_7$ (bottom) and Y$_3$NbO$_7$ (top) and the one obtained from a
perfect fluorite structure (dashed line) at room temperature. The experimental data
(dots) from \cite{Norb} obtained via bond valance sum constrained RMC are shown as well.}}
\label{RMCvsMD}
\end{figure}

The disordered character of the anion sublattice can be appreciated from the
oxygen-oxygen radial distribution functions taken from the room temperature runs,
$g_{\rm{O-O}}(r)$ (see figure \ref{RMCvsMD}). They are compared with those expected from
a perfect fluorite structure with the same lattice parameter (and with the
$g_{\rm{O-O}}(r)$ obtained from the RMC analysis of the experimental data \cite{Norb}).
The first peak in $g_{\rm{O-O}}(r)$ for Y$_3$NbO$_7$ is shifted by 0.13~\AA\ from the
expected position for an ideal fluorite structure and the second peak (which
indicates the distance between two next-nearest neighboring oxygen ions) is
broadened and presents a number of subsidiary features. Zr$_2$Y$_2$O$_7$ shows
similar but less marked deviations from the ideal fluorite order, as already noted
from the diffuse scattering intensity \cite{Norb,Lee1}. \newline

\begin{figure}[htbp]
\begin{center}
\includegraphics[width=8cm]{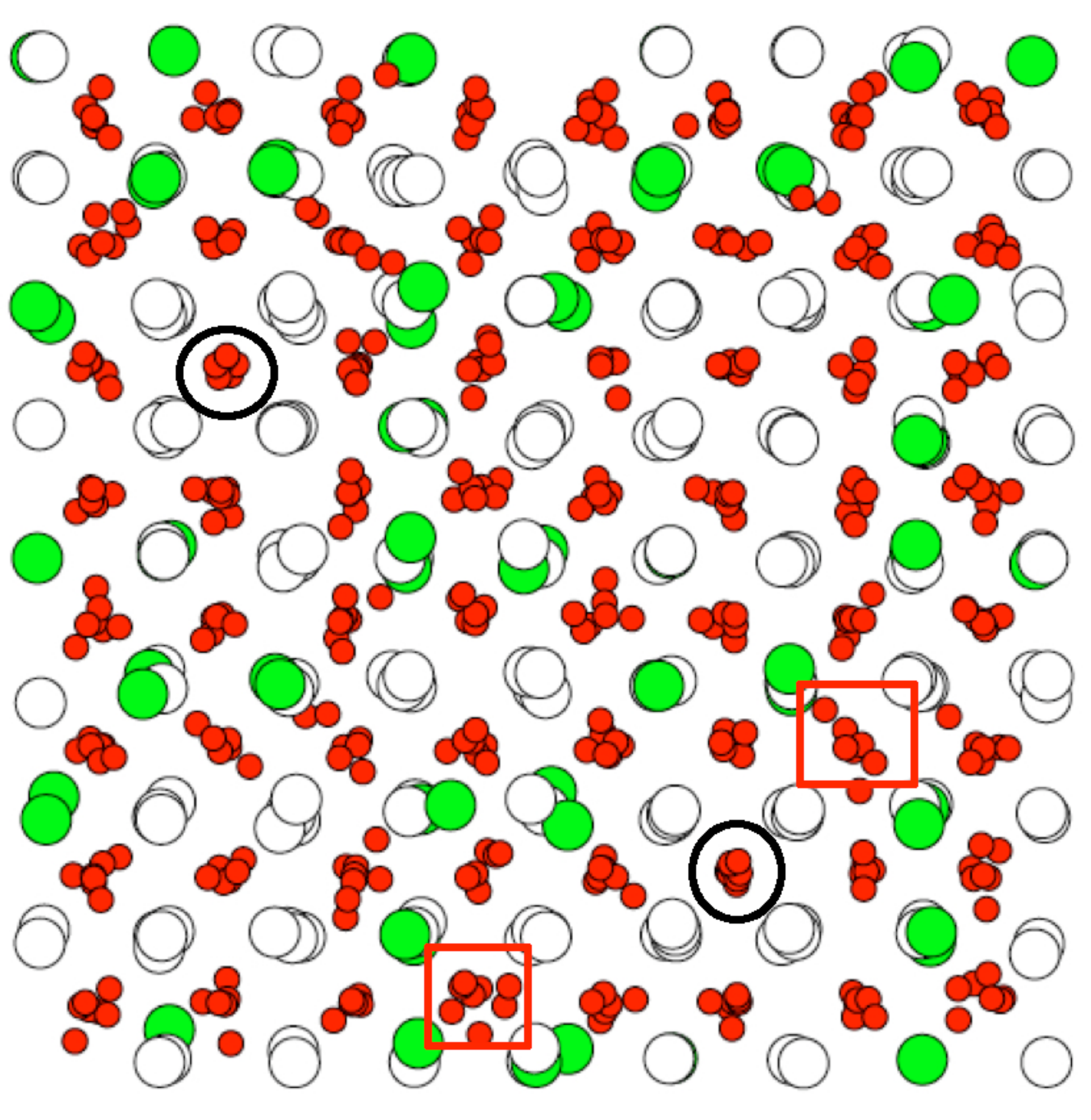}
\end{center}
\caption{\it{\small Snapshot along the z axis of the simulation  box in
Y$_3$NbO$_7$. The color code is white, green and red for Yttrium, Niobium and
Oxygen ions respectively. (In black and white, the colour code is white,
light grey and dark grey for Yttrium, Niobium and Oxygen ions respectively.)
The two black circles show those oxygen ions which are
surrounded by Yttrium ions only, whilst the two red squares show those which are
surrounded by both Niobium and Yttrium ions. }} \label{Snapshot}
\end{figure}

In the fluorite structure, the anions sit in a tetrahedral site.  Because the
cation species are distributed randomly in the {\it d-}fluorite Zr$_2$Y$_2$O$_7$-Y$_3$NbO$_7$
system, the vertices of these tetrahedra may be occupied by different combinations
of cations. A snapshot of the simulation cell in Y$_3$NbO$_7$ can be seen in figure
\ref{Snapshot}. The two black circles highlight those oxide ions that are surrounded by Y$^{3+}$
cations only and the red squares those with a significant number of Nb$^{5+}$ ions
in their coordination environment. Notice that the former are much more tightly
grouped about the lattice site than the latter. The notion that the Nb$^{5+}$ cations are
somehow responsible for the disorder observed in Y$_3$NbO$_7$ is confirmed by the
fact that the Y-O and Nb-O distances, obtained from the RMC analysis of the neutron
scattering data in reference \cite{Norb} (or analogously from our simulations), are quite different ($d_{\rm{Nb-O}}$ = 1.96 \AA\,
$d_{\rm{Y-O}} $= 2.28 \AA). In a perfect fluorite structure, the anion-cation distance
is $d_{an-cat}$ = $\sqrt{3}/4$ $a$ = 2.27 \AA, where $a$ is the lattice parameter obtained 
from the RMC analysis. It is therefore evident that the Nb$^{+5}$
ions are attracting the oxygen ions very tightly, thus disrupting the perfect
fluorite order.  This effect is less significant in Zr$_2$Y$_2$O$_7$ because the
cation charges are 3+ and 4+ in this case and because the cation radii are more
similar. \newline

\newpage

\subsection{Cation effects on anion mobility}
\begin{figure}[htbp]
\begin{center}
\includegraphics[width=8cm]{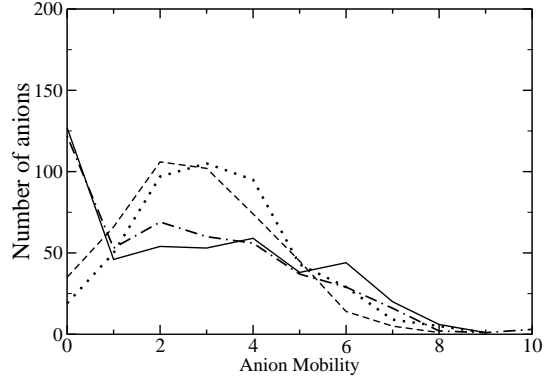}
\end{center}
\caption{\it{\small Number of ions with certain mobility against  the mobility
itself for Y$_3$NbO$_7$ (solid line), Zr$_2$Y$_2$O$_7$ (dashed line), equal-charge Y$_3$NbO$_7$ (dotted line) and
equal-radius Y$_3$NbO$_7$ (dot-dashed line).}}
\label{Hoppers_distrib}
\end{figure}

The inhomogeneities induced by this strain lead to a significant difference in
conduction mechanisms between Zr$_2$Y$_2$O$_7$ and Y$_3$NbO$_7$. In figure
\ref{Hoppers_distrib} we  show the probability distribution (calculated as in \cite{Norb}) for the individual
oxide ions to make a certain number of diffusive hops (by the O-O nearest-neighbour
separation) during the course of the simulations at 1500 K. In Y$_3$NbO$_7$ a large
fraction of anions do not hop at all during the whole simulation whereas a few
anions possess a high mobility. The former make no contribution to the
conductivity. Zr$_2$Y$_2$O$_7$, on the other hand, has a more homogeneous behaviour
with fewer immobile oxygen ions and a probability distribution which peaks about
the average mobility of 2-3, with relatively few outliers. Further examination (see
below) shows that the immobile oxide ions in Y$_3$NbO$_7$ are in tetrahedral sites
where all the vertices are occupied by Y$^{3+}$ cations. \newline

As a consequence, the conductivity in Y$_3$NbO$_7$ exhibits an ``aging"
effect, especially at low temperatures. The average mobility shortly after
initialising the simulation with randomly placed oxide ions is high, but drops over
a substantial period of time ($\sim$1 ns) presumably because, on this time scale,
oxide ions find initially empty Y$^{3+}$-rich sites and become trapped there.
Another consequence is that different initial configurations give different mean
diffusion coefficients even after very long runs. This is probably caused by the
variation in the numbers of different types of tetrahedral sites in the
initialisations. Both of these effects make a reliable determination of the
conductivity difficult for Y$_3$NbO$_7$ in the temperature range ($<1500$ K) where
experimental measurements have been made. Our best efforts give a conductivity
($\sigma_{NE} = 0.016 \;\Omega^{-1} \rm{cm^{-1}}$ at 1500 K, where the subscript NE
denotes Nernst--Einstein, see below) of about an order of
magnitude larger than measured \cite{Norb,Lee1}. This value is much lower than the
one obtained for Zr$_2$Y$_2$O$_7$ ($\sigma_{NE} = 0.08 \;\Omega^{-1} \rm{cm^{-1}}$),
which agrees with the experimental trend \cite{Norb}. For Zr$_2$Y$_2$O$_7$ neither
of these effects is observed and reliable conductivity values can be obtained down
to the highest temperatures at which measurements are available \footnote{There is
a lower limit to the values of the diffusion coefficient and conductivity which may
be determined from the mean-squared displacement of the ions as a function of time.
To be reliable, the ions must on average be diffusing by more than a lattice
parameter, and when the mobility is very low this requires the mean-square
displacement to be measured over a very long time and eventually longer than we can
afford to simulate.}. In this case the slope of conductivity {\it vs.} inverse
temperature agrees very well with experiment and the actual value at 1500 K is only
slightly larger than experiment (by a factor of less than 2 for the data in \cite{Lee1}). \newline

The above-mentioned conductivity values were calculated from the ionic diffusion
coefficients by using the Nernst--Einstein formula
 \begin{equation}
    \sigma_{NE} = \frac{c^2\rho D}{k_BT}
    \label{Nernst-Einstein}
    \end{equation}
where $c$ is the charge, $\rho$ the number density and $D$ the diffusion
coefficient of the mobile species, $k_B$ is the Boltzmann constant and $T$ is the
temperature. The diffusion coefficients themselves have been calculated from the
long-time slopes of a plot of the mean-squared displacement (msd) of individual
ions versus time. The Nernst--Einstein formula assumes that the ions move
independently, which can only be approximately true in reality. A better value for
the conductivity is obtained, in principle, from an integral over the charge
current correlation function $J(t)$\cite{Hansen}:
\begin{equation}
\sigma=\frac{1}{k_B T V} \int_0^{\infty} J(t) dt
 \label{true_conduct}
\end{equation}
though this quantity has very poor statistics, especially when diffusion is slow.
We were able to get a reliable value of the true conductivity from very long runs
on the most highly conducting Zr$_2$Y$_2$O$_7$ system. At 1500 K
$\sigma=0.05\;\Omega^{-1} \rm{cm^{-1}}$ which is 0.6 times smaller than the
Nernst-Einstein value and in much better agreement with the experimental values in
\cite{Norb, Lee1}. At higher temperatures the differences between the two methods of
calculating the conductivity become smaller.

\section{The effects of cation charges and sizes on the material's properties}

The simulations on the Zr$_{0.5-0.5x}$Y$_{0.5+0.25x}$Nb$_{0.25x}$O$_{1.75}$  series
in the {\it d-}fluorite structure give the same trend of decreasing conductivity
and increasing disorder with increasing $x$ that is seen experimentally. In this
section we will attempt to distinguish the effects of cation size and charge by
comparing the results obtained with modified interaction potentials with the
realistic models.

\subsection{The effect of the cation charge} \label{charge}
In order to examine the role of the high charge of the Nb$^{5+}$ cation in causing
these effects we have carried out simulations on the $x$=1 system in which both the
``Nb" and ``Y" ions are assigned a charge of 3.5+ (which maintains charge
neutrality for this stoichiometry). We will call this system {\it equal-charge}
Y$_3$NbO$_7$. All other aspects of the interaction potentials were kept the same
(see \ref{Sim-det}) as in the runs described above. We started from a
previous high temperature simulation on Y$_3$NbO$_7$ and equalized the cation
charges. All the simulations were equilibrated at constant pressure for 100,000
steps, during which the velocities were rescaled several times to keep the system
at the required (high) temperatures for conductivity studies, and then longer
(constant volume) runs were made to study the properties of this material. We also
ran some shorter simulations at room temperature to study the static structure and
obtain the radial distribution functions. \newline

\begin{figure}[htbp]
\begin{center}
\includegraphics[width=8cm]{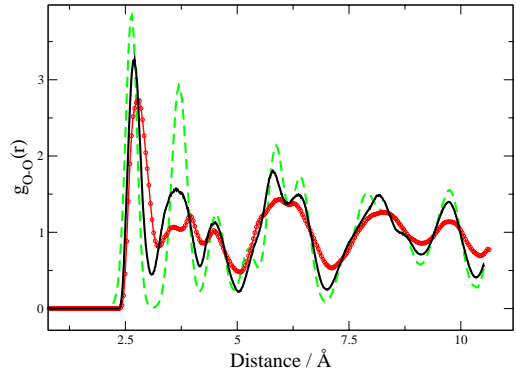}
\end{center}
\caption{\it{\small Comparison between the oxygen-oxygen g(r) for Y$_3$NbO$_7$ (red
dotted line), equal-charge Y$_3$NbO$_7$ (black solid line)  and a perfect fluorite
structure (red dashed line).}} \label{RDF11}
\end{figure}

In figure \ref{RDF11} we show the oxygen-oxygen radial distribution functions for
room temperature {\it equal-charge} Y$_3$NbO$_7$ , Y$_3$NbO$_7$ itself,  and a
perfect fluorite structure with a matched lattice constant. It is clear that
equalizing the cation charges increases the order of the oxide ion sublattice 
\footnote{A movie can be found in the online version of this paper which shows
how Y$_3$NbO$_7$ becomes more ordered once the cation charges are equalized.}. The
$g_{\rm{O-O}}(r)$ for the equal-charge system shows a first peak close to the ideal
fluorite position and a single prominent second peak, though still not as intense
as in the ideal case. We also find that the cation-anion distances change
significantly, $d_{\rm{Y-O}}=2.23$ \AA\ and $d_{\rm{Nb-O}}=2.19$ \AA . This is another
sign that the system has become more ordered, as these become closer to the ideal
cation-anion distance in this system, given, in this case, by $\sqrt{3}/4$ $a_{equal-charge}$ = 2.27 \AA. 
A comparison between figure \ref{RDF11} and figure \ref{RMCvsMD} shows that {\it
equal-charge} Y$_3$NbO$_7$ is also more ordered than Zr$_2$Y$_2$O$_7$ which is
consistent with the fact that Zr$_2$Y$_2$O$_7$ still has two cations with different
charges (4+ and 3+).\newline

Figure \ref{Hoppers_distrib} shows that equalizing the cation charges (dotted line)
greatly reduces the number of immobile oxide ions, relative to Y$_3$NbO$_7$. In
fact, the distribution comes to resemble that for Zr$_2$Y$_2$O$_7$ quite closely
and is indicative of a homogeneous pattern of diffusion. The mean  squared
displacement for the oxide ions in {\it equal-charge} Y$_3$NbO$_7$ was calculated
at $1500$ K and $2000$ K and the respective diffusion coefficients are reported in
table \ref{DiffCoeff}. From this it can be seen that {\it equal-charge}
Y$_3$NbO$_7$ is more conducting than Y$_3$NbO$_7$. Interestingly {\it equal-charge}
Y$_3$NbO$_7$  has a lower diffusion coefficient than Zr$_2$Y$_2$O$_7$, 
whereas we have just seen that the latter system has a more
disordered oxide ion lattice. This suggests that there is some other factor at work
besides the charge-induced strain in the lattice. \newline

\begin{figure}[htbp]
\begin{center}
\includegraphics[width=8cm]{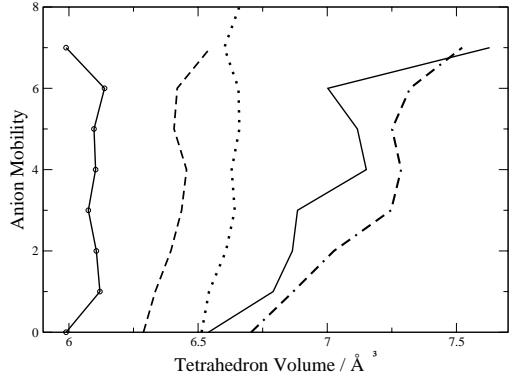}
\end{center}
\caption{\it{\small Anion mobility as a function of the average tetrahedron  volume
for {\it equal-charge} (dashed line), equal-radius (dot-dashed line), {\it
equal-charge}\&radius Y$_3$NbO$_7$ (circles+line), Y$_3$NbO$_7$ (solid line) and
Zr$_2$Y$_2$O$_7$ (dots). }} \label{MobVSTetVol}
\end{figure}

\begin{table}
\caption{\label{DiffCoeff} Diffusion coefficients.}
\begin{indented}
\item[]\begin{tabular}{@{}lllll}
\br
Material            &   $10^{6}\times D_O $  (\rm{cm $^2$} $s^{-1}$)   &\\
\mr
Y$_3$NbO$_7$ & 0.924    &\\
Equal-charge Y$_3$NbO$_7$ & 1.87  &  $T=2000$ K\\
Zr$_2$Y$_2$O$_7$ &   2.87  &\\
\br
Y$_3$NbO$_7$ & 0.072    &\\
Equal-charge Y$_3$NbO$_7$ & 0.21  &  \\
Zr$_2$Y$_2$O$_7$ &    0.44 & $T=1500$ K\\
Equal-radius Y$_3$NbO$_7$ & 0.37 &   \\
Equal-charge\&radius Y$_3$NbO$_7$ & 2.6 &   \\
\br
\end{tabular}
\end{indented}
\end{table}

In order to see {\em how} the dopant cation charge induces the disorder and affects
the ionic mobility we calculated the mobility distribution as done in \cite{Norb}
but this time we also monitored the identities of the four cations
surrounding the anion. As already anticipated above, the anions with no mobility
(about 1/4 of the total number, see figure \ref{Hoppers_distrib}) are in sites with
a much higher Y$^{3+}$ content than expected ($\approx$ 3.65 as opposed to the
average of 3 Y$^{3+}$ and 1 Nb$^{5+}$ which is expected from the Y$_3$NbO$_7$
stoichiometry) and the higher the mobility the higher the content of Nb$^{5+}$ ions
in the tetrahedron surrounding the anion. This effect seems to be associated with
the {\em volume} of the coordination tetrahedron (obtained from the volume of the
tetrahedron with the four closest cations at its vertices). The Coulombic repulsion
between two Nb$^{5+}$ ions is almost three times stronger than that between the
Y$^{3+}$ cations and from the cation cation radial distribution functions we find
that the nearest-neighbour cation-cation separations are $d_{\rm{Y-Y}}$ = 3.69 \AA\ and
$d_{\rm{Nb-Nb}}$ = 3.86 \AA\ (the ideal fluorite value being $a/ \sqrt(2)=3.78$ \AA) so
that the coordination tetrahedra are strained to an extent which depends on their
Nb$^{5+}$ content with the Nb$^{5+}$-rich ones significantly larger than the
average. The anion mobility is plotted against the average volume of the
surrounding tetrahedra in figure \ref{MobVSTetVol} (solid line). Alternatively, we
could have plotted the mobility against the average number of Nb$^{5+}$ ions in the
coordination shell and obtained a very similar curve. Note, firstly, that the
tetrahedral volumes for Y$_3$NbO$_7$ span an appreciable range, from 6.4 to 7.5
\AA$^3$ and secondly, that the anion mobility is a strongly increasing function of
the tetrahedron volume. This observation therefore explains the paradox that
although highly charged Nb$^{5+}$ cations bind oxide ions strongly they promote
{\em local} mobility but reduce the overall conductivity by straining the cation
sub-lattice resulting in the formation of oxide ion trapping sites. We also carried
out the same analysis for Zr$_2$Y$_2$O$_7$ (dotted curve) and for {\it
equal-charge} Y$_3$NbO$_7$ (dashed curve) and present the results in figure
\ref{MobVSTetVol}. In these cases, the range of tetrahedron volumes is much
narrower than for Y$_3$NbO$_7$ ($<0.2$ \AA$^3$) suggesting a much less strained
cation lattice and the dependence of the anion mobility on the tetrahedral volume
is barely discernable.

\subsection{The effect of the cation size} \label{cation-radius}
\begin{figure}[htbp]
\begin{center}
\includegraphics[width=8cm]{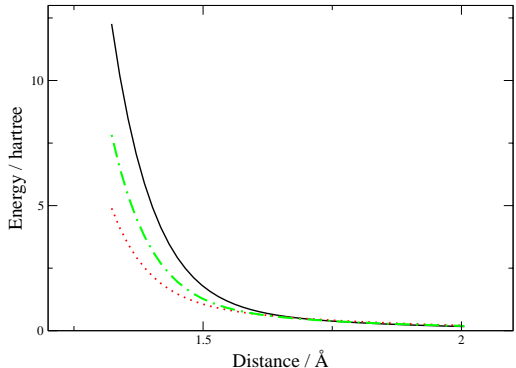}
\end{center}
\caption{\it{\small Short-range repulsion parts of the potential for the Y-O (solid line), Nb-O (dotted line) and Zr-O (dot-dashed line) terms. }}
\label{SR_potential}
\end{figure}

Proceeding by analogy with the {\it equal-charge} study, we can examine the effect of
equalizing the cation radii in Y$_3$NbO$_7$. The short-range repulsion terms in our
potentials model the repulsion between two ions and therefore are intrinsically
related to the extent of the valence electron density around the ion. In figure
\ref{SR_potential} we show the short-range repulsion terms for the Y-O, Nb-O and
Zr-O interactions (recall that these were generated with the {\it ab initio}
force-fitting strategy). It can be seen that the range of these potentials is
consistent with the fact that Y$^{3+}$ is the largest ion, Nb$^{5+}$ the smallest,
with Zr$^{4+}$ of intermediate size. To generate an equal-radius model for
Y$_3$NbO$_7$
we used the Zr-O short-range interaction parameters for the Y-O and Nb-O terms,
but the Y$^{3+}$ and Nb$^{5+}$ ions keep their formal charges \footnote {Note that
the Y-O and Nb-O nearest-neighbour separations which result from these simulations
will differ, because of the stronger Coulombic attraction to the Nb$^{5+}$ ion.
From an empirical perspective, {\it i.e.} the way that the ionic radius is normally
assigned, the ions do not appear as of equal size; rather they have been treated as
if the electron density distributions around the ions are of equal size in the
potential model.}. Simulations are run as previously described for the equal-charge
case.
\newline

The partial radial distribution functions for {\it equal-radius} Y$_3$NbO$_7$ looks
quite similar to the one obtained for Y$_3$NbO$_7$ itself (figure \ref{RMCvsMD} and
reference \cite{Norb}) and we have not shown them separately. Despite the equal
radius potential construction, even  the anion-cation distances are comparable
($d_{\rm{Nb-O}}=1.93$ \AA, $d_{\rm{Y-O}}=2.29$ \AA) to those of the parent compound. This
confirms that the observed disorder is caused by the different cation charges in
Y$_3$NbO$_7$ and that the different ranges of the repulsive cation-anion
interactions do not play an important role. Given this similarity in structure, it
is not surprising that the  equal-radius system shows a wide range of tetrahedral
volumes like Y$_3$NbO$_7$ itself and figure \ref{MobVSTetVol} (dot-dashed line)
shows that, again, the oxide ion mobility tracks the tetrahedral volume like the
parent compound. This leads to the inhomogeneous distribution of hopping
probabilities, as shown in figure \ref{Hoppers_distrib} (dot-dashed line) with a
substantial fraction of immobile ions.
\newline

In Table \ref{DiffCoeff} we report the average diffusion coefficient obtained  from
a simulation at $1500$ K on {\it equal-radius} Y$_3$NbO$_7$ and compare this value
with {\it equal-charge} Y$_3$NbO$_7$, Y$_3$NbO$_7$ and Zr$_2$Y$_2$O$_7$ at the same
temperature. Interestingly, the equal-radius material is more conducting than both
{\it equal-charge} Y$_3$NbO$_7$ and Y$_3$NbO$_7$ though not as conducting as
Zr$_2$Y$_2$O$_7$. That {\it equal-radius} Y$_3$NbO$_7$ is more conducting than
Y$_3$NbO$_7$ itself is probably to be expected, what is more surprising is that it
is also more conducting the {\it equal-charge} system which, as we have seen, has
the kind of homogeneous distribution of mobilities seen in the highly conducting
Zr$_2$Y$_2$O$_7$ system. This is probably a consequence of using the Zr-O short-range interaction
parameters for the Y-O and Nb-O terms. As we mentioned above, these are softer than the ones for the
Y-O interaction but harder than the Nb-O ones. However, since in this system there are three times more Y$^{3+}$ ions than Nb$^{5+}$ ions,
this means that, on average, the anion-cation short-range repulsion in this system is softer and
this greatly enhances the anion mobility. A similar phenomenon was observed in the case of GeO$_2$ where
rescaling the short-range interaction terms by a factor of 1.6 enhances the oxide-ion diffusivity
by several orders of magnitude \cite{Marrocchelli2009}. \newline

As a final model system, we equalize both the cation radii and charges
simultaneously ({\it i.e.} we use the Zr-O potential for all short-range
interactions and set all cation charges equal to 3.5. As might be expected this
gives a small distribution of tetrahedral volume sizes (figure \ref{MobVSTetVol},
line with circles) and virtually no oxide ion trapping. This material is an incredibly good
ionic conductor with a conductivity of $\sigma_{NE} \approx \; 0.6\; \Omega^{-1}\rm{cm^{-1}}$
at $1500$ K, a value higher than the one found in the best yttria-stabilized
zirconias ($\sigma \approx \; 0.6\; \Omega^{-1}\rm{cm^{-1}}$ at $1667$ K for 8\% YSZ \cite{Subbarao}).
\newline

\section{Vacancy ordering effects}

So far we have examined the relationship between disorder and mobility by examining
the ionic positions and the effect of making particular changes in the interaction
potential. In the Introduction, we motivated our study of this system by reference
to the properties of the {\em vacancies} and the way they are ordered by their
interactions with the ions. Their effect is, of course, implicit in the properties
we have examined so far, but now we turn to an examination of the simulation
trajectories by considering the vacancies explicitly.

\subsection{Vacancy interactions}
\begin{figure}[htbp]
\begin{center}
\includegraphics[width=8cm]{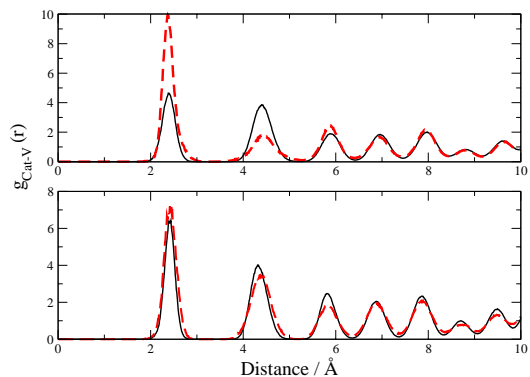}
\end{center}
\caption{\it{\small Yttrium-Vacancy (solid line) and Niobium-Vacancy (dashed line)
radial distribution functions in {\it d-}fluorite structured Y$_3$NbO$_7$ (top) and equal-charge Y$_3$NbO$_7$ (bottom).}}
\label{Cat-Vac_RDF}
\end{figure}

In \ref{Vacancy} we explain how we identify the vacancies and how we can obtain the
ion-vacancy and vacancy-vacancy radial distribution functions. Integration of these
can be used to define coordination numbers. For example, integrating the
Y$^{3+}$-vacancy rdf, $g_{\rm{Y-V}}(r)$, from zero out to the position, $r_c$, of first
minimum of the $g_{\rm{Y-O}}(r)$ gives the average number of vacancies in the first
coordination shell of a ${\rm Y}^{3+}$ ion. In figure \ref{Cat-Vac_RDF} we show the
cation-vacancy radial distribution functions for Y$_3$NbO$_7$ and {\it
equal-charge} Y$_3$NbO$_7$ at 1500 K. From this figure the tendency of the
vacancies to bind to Nb$^{5+}$ ions is clear for Y$_3$NbO$_7$. This tendency is
strongly reduced when the cation charges are equalized. \newline

In table \ref{Cat-Vac} we report the number of vacancies around the different
cations compared to the number which would be expected if the vacancies occupied
sites randomly. It is clear from the table that vacancies
bind very strongly to Nb$^{5+}$ ions in {\it d-}fluorite Y$_3$NbO$_7$ and less
strongly to Zr$^{4+}$ in Zr$_2$Y$_2$O$_7$. Considering the Coulombic effects, the
vacancies (which have a relative charge of 2+) would be expected to bind more
strongly to the cation with the lowest charge, {\it i.e.} Y$^{3+}$ in both materials. The
fact that the contrary happens is therefore indicative of the importance of strain
effects which are strong enough to outweigh the Coulombic interactions \cite{Bog1},
{\it i.e.} the vacancies prefer to bind to the smallest cations, despite their
higher charges. This is confirmed by solid state NMR studies \cite{Kawata2006} on 
yttria-doped zirconias. \newline

\begin{table}
\caption{\label{Cat-Vac} Cation-Vacancy relative coordination numbers extracted from our MD simulations at 1500 K.}
\begin{indented}
\item[]\begin{tabular}{@{}lllll}
\br
Material & $n_{Y-V}$ & $n_{Nb-V}$ & $n_{Zr-V}$ \\
\mr
RANDOM & 1 & 1 & 1\\
Y$_3$NbO$_7$ & 0.81 & 1.53 & \\
Zr$_2$Y$_2$O$_7$ & 0.79 & & 1.21\\
Equal-charge Y$_3$NbO$_7$  & 0.92 & 1.23 & \\
Partially ordered Y$_3$NbO$_7$  & 0.72 & 1.85 & \\
Pyrochlore Zr$_2$Y$_2$O$_7$  & 0.15 & & 1.85  \\
\br
\end{tabular}
\end{indented}
\end{table}

We may now extend this analysis to examine the vacancy-vacancy ordering by
integrating $g_{\rm{V-V}}$ between limits which bound the different coordination shells
around an anion site in an ideal fluorite lattice. The first such shell gives 
the probability of finding a pair of vacancies as nearest-neighbours ({\it
i.e.} along the $\langle 100\rangle$ direction of the fluorite lattice) whereas the
second shell relates to vacancy pairs along $\langle110\rangle$ and the third to
$\langle111\rangle$. These probabilities may be compared with what would be
expected if the vacancies are randomly distributed on the simple cubic anion
lattice and we give the ratios in table \ref{Table III} for simulations at 1500 K.
As we note in the \ref{Vacancy}, it is likely that the way that we identify
vacancies leads to an overestimate of the number of vacancy nearest-neighbours.
Nevertheless, it can be seen that there is a specific ordering tendency, with
$\langle111\rangle$ vacancy pairs appearing with a higher frequency than would be
expected for a random distribution, and nearest-neighbour pairs being lower. This
order is in accord with Bogicevic's \cite{Bog1} examination of the energetics of
vacancy ordering in lightly doped yttria-stabilised zirconia. The tendency we
observe is, however, not as strong as might be expected from the calculations in
this paper. This is, in part, due to the above-mentioned overestimate of the number
of nearest-neighbours for algorithmic reasons, but is also because our calculations
have been performed at high temperatures where the ordering is opposed by entropic
effects. It is also interesting to notice that the vacancy ordering tendency
decreases as temperature is raised and completely disappears at 2000 K. Finally,
this tendency seems to be stronger in Zr$_2$Y$_2$O$_7$ than in Y$_3$NbO$_7$ which
might seem at first counterintuitive considering that Zr$_2$Y$_2$O$_7$ is a better
conductor than Y$_3$NbO$_7$. This apparent contradiction will be explained in the
next section.

\begin{table}
\caption{\label{Table III} Relative population of the anion vacancy pairs expected
from a random distribution of vacancies within a cubic fluorite lattice, compared
with those extracted from our MD simulations at $1500$ K.}
\begin{indented}
\item[]\begin{tabular}{@{}lllll}
\br
Material            &   $\left <100\right >$  &  $\left <110\right >$ &  $\left <111\right >$ \\
\mr
RANDOM & 0.231 & 0.461 & 0.308 \\
Y$_3$NbO$_7$ & 0.191 & 0.418 & 0.390\\
Zr$_2$Y$_2$O$_7$ & 0.166 & 0.391 & 0.442 \\
Equal-charge Y$_3$NbO$_7$  & 0.136 & 0.480 & 0.384 \\
Partially ordered Y$_3$NbO$_7$  & 0.154 & 0.304 & 0.542\\
Pyrochlore Zr$_2$Y$_2$O$_7$  & 0.073 & 0.147 & 0.780  \\
\br
\end{tabular}
\end{indented}
\end{table}

\subsection{Effects of vacancy interactions on the diffuse scattering}
The vacancy ordering influences the diffuse scattering observed in diffraction
studies, and it is useful to compare the consequences of the degree of vacancy
ordering we find in the simulations with those seen experimentally. Our simulations
have reproduced extremely well the structures derived from the analysis of the
total ({\it i.e.} Bragg plus diffuse) {\em powder} diffraction \cite{Norb} in these
systems. However, several SAED studies \cite{Miida, Irvine2000} have shown
intense diffuse peaks for Y$_3$NbO$_7$ at particular positions in reciprocal space.
This suggests that the vacancy ordering effects are considerably stronger and of
longer range than is indicated by the propensities we have shown in tables
\ref{Cat-Vac} and \ref{Table III}. In order to compare with these experimental
studies we have calculated the diffuse scattering from the ionic configurations
available from the simulations as described in \ref{Diffuse}.
\newline

 We decided to calculate the electron diffraction pattern of Y$_3$NbO$_7$ and
Zr$_2$Y$_2$O$_7$ along the [\=1\=10]$_f$ zone axis ($f$ refers to the cubic fluorite
structure). This is found to be the most informative pattern in the experimental
studies \cite{Miida,Irvine2000,Whittle}. In the experimental patterns of Y$_3$NbO$_7$,
a pair of intense and closely-spaced diffuse peaks is seen at ${\bf G_F}\pm \frac{1}{2}(111)^{*}$, where
${\bf G_F}$ is a reciprocal lattice vector of the fluorite-type lattice. In fact,
the two peaks correspond to a ring of diffuse scattering centred at these positions
which is projected down into the [\=1\=10]$_f$ plane. The ${\bf G_F}\pm
\frac{1}{2}(111)^{*}$ positions correspond to the location of Bragg peaks in the
pyrochlore structure and the diffuse scattering seen in Y$_3$NbO$_7$ indicates a
pattern of vacancy pairs along the [111]$_f$ direction which form strings along
[110]$_f$, as in the pyrochlore structure, but of finite range. \footnote{An ideal
pyrochlore structure type, $A_2$$B_2$O$_7$,  is a superstructure of the fluorite
($M$X$_2$) structure and is based upon a 2 x 2 x 2 unit cell with Fd\={3}m symmetry
(figures showing the differences between a fluorite and a pyrochlore structure can
be found in refs \cite{SHu1} \cite{Whittle}.) The cation superstructure of
pyrochlore is based upon ordering of $A$ and $B$ cations parallel to the
$\langle110\rangle$ directions, separated by $\frac{1}{2} \frac{1}{2} \frac{1}{2}$,
with respect to the origin. The $A$ and $B$ cations are, respectively, found at the 16c
(eight-coordinated) and 16d (six-coordinated) sites whereas the anions are
distributed between two tetrahedrally coordinated positions: 48f [O(1)] and 8a
[O(2)]. There is another tetrahedral site potentially available for the anions, 8b,
which is systematically vacant in fully ordered pyrochlores.} In Zr$_2$Y$_2$O$_7$
the diffuse scattering is considerably weaker and the pattern of peaks is somewhat
different. It appears as streaks along [001]$_f$ in the studies by Irvine {\it et
al} \cite{Irvine2000} and taken as indicative of C-phase ordering of vacancies
({\it i.e.} vacancy pairs along [110]$_f$). Whittle {\it et al} \cite{Whittle},
however, observe the pattern as 3 peaks centred at ${\bf G_F}\pm
\frac{1}{2}(111)^*$ and interpret it as another type of pyrochlore-like modulated
structure. \newline

\begin{figure}[htbp]
\begin{center}
\includegraphics[width=8cm]{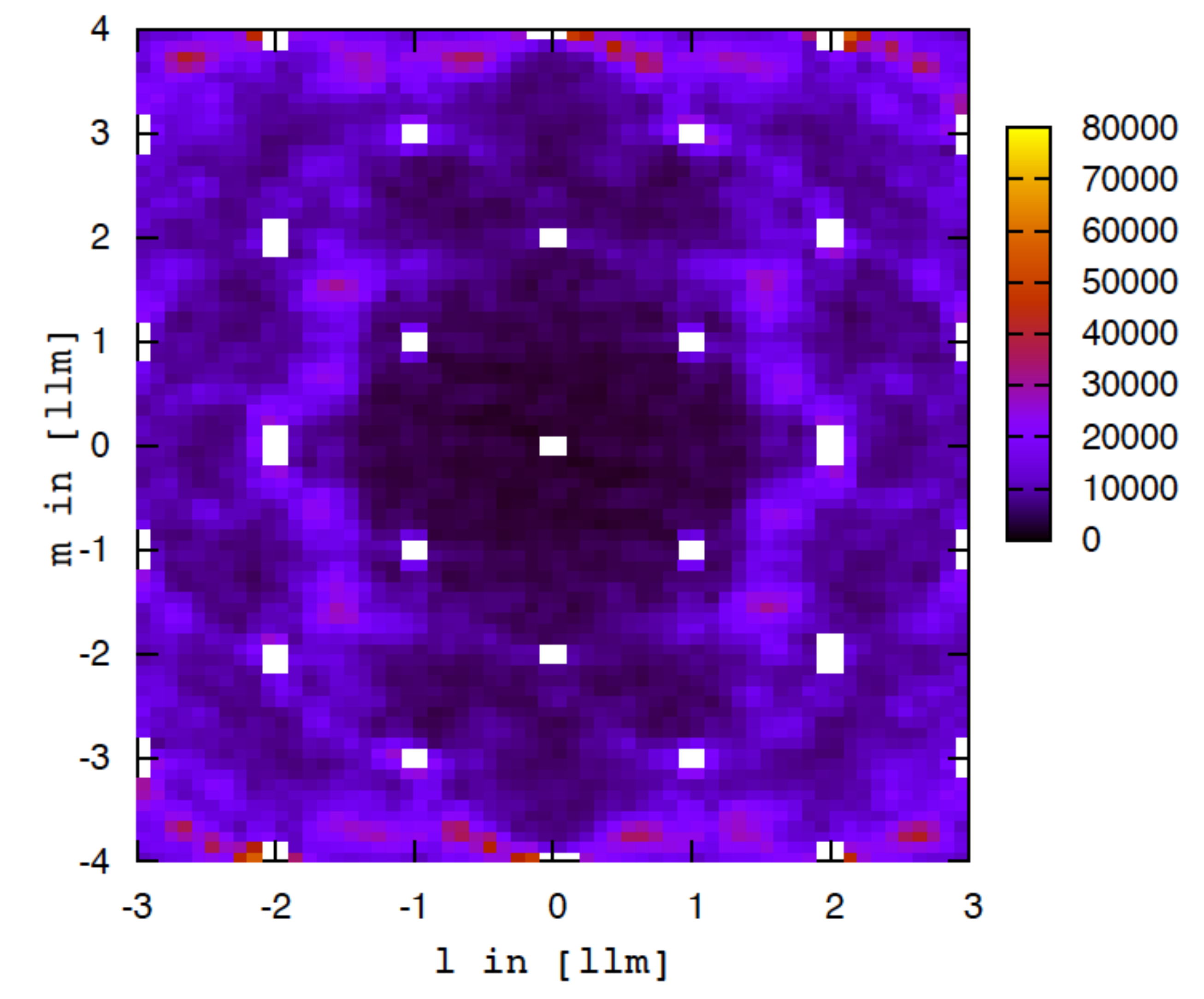}
\includegraphics[width=8cm]{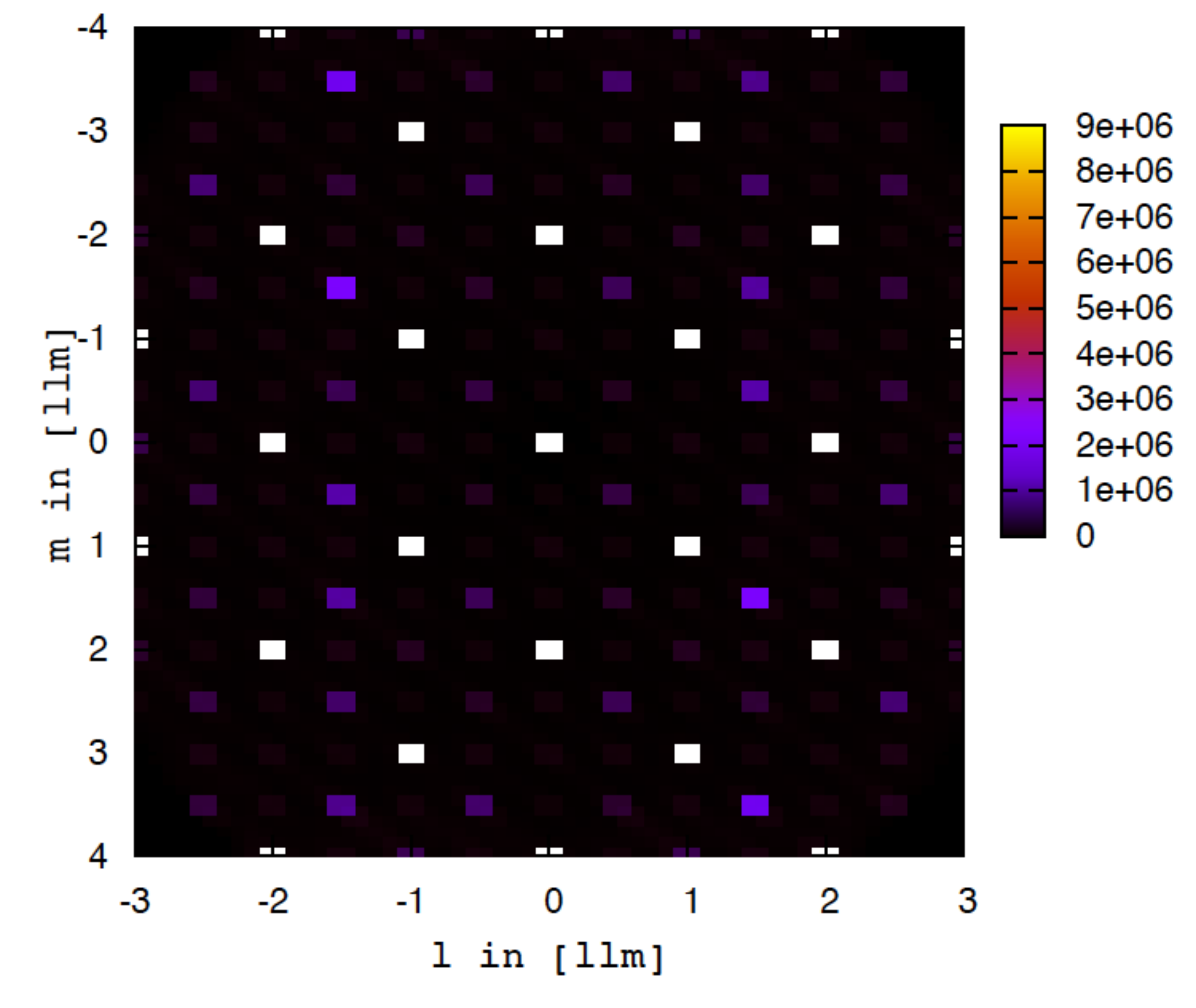}
\end{center}
\caption{\it{\small Electron diffraction pattern obtained from the MD data for d-fluorite (top) and locally cation-ordered (bottom) Y$_3$NbO$_7$.}} \label{ED1}
\end{figure}

In figure \ref{ED1} we show the electron diffraction patterns calculated from our
MD data (obtained from a simulation on 11,000 atoms, {\it i.e.} 10 x 10 x 10 unit cells).
The pattern does not show the diffuse spots at ${\bf G_F}\pm \frac{1}{2}(111)^*$
which have been observed in the experimental SAED patterns \cite{Miida,Irvine2000,
Whittle}. The calculated patterns show some diffuse signal at some of these positions but 
this appears weaker and less sharply peaked than in the experimental studies. This finding was
confirmed in longer MD simulations on smaller cells. As a test, we made a similar
comparison with the diffuse neutron scattering \cite{Hull1999} derived from
single-crystal studies of (ZrO$_2$)$_{1-x}$-(Y$_2$O$_3$)$_x$ for $0.1<x<0.25$ and
this was very successful, thus indicating that the disagreement with the
experimental data is probably not a limitation of the way the diffuse scattering
has been calculated.
\newline

For Y$_3$NbO$_7$ we prepared a starting configuration for an MD run in which the
cations were distributed randomly over the cation sublattice but in which the oxide ions
were placed as if in a pyrochlore structure, to give the associated vacancy-ordered
structure. In simulations at 1500 K, this initial configuration rapidly evolved to
one in which the extent of vacancy ordering had relaxed to that seen in the
previous {\it d-}fluorite simulations. It therefore does not appear that we can
reproduce the evidence for vacancy ordering from the SAED studies with simulations
based upon the {\it d-}fluorite structure with randomly disposed cations. \newline

In summary then, the {\it d-}fluorite simulations reproduce the experimentally
observed {\em trends} in disorder and conductivity and have enabled us to account
for the way that these properties are affected by the cation size and charge.
However, there is a suggestion that these simulations overestimate the
conductivities, especially on the Nb-rich side of the composition range.
Furthermore, although the simulations demonstrate vacancy ordering tendencies
similar to those which have been found in more lightly doped zirconias these are
not sufficiently strong to reproduce the SAED patterns. We are therefore led to
reconsider one of the underlying assumptions of the work described so far, namely
that the different cation species are randomly distributed throughout the lattice.

\section{Local cation ordering effects} \label{cat-ordering}

Our simulation studies of Y$_3$NbO$_7$ in the {\it d-}fluorite structure have shown
lattice distortions induced by the Nb$^{5+}$ ions, especially when two or more
Nb$^{5+}$ ions are found as cation nearest-neighbours (the nearest neighbour
Nb$^{5+}$-Nb$^{5+}$ distance is 3.86 \AA\ as opposed to the ideal cation-cation
distance in a fluorite structure $a/ \sqrt(2)=3.78$\AA). The strength of this
effect would suggest that the system will try to adapt by ordering the Nb$^{5+}$
ions so as to minimise the strain. Furthermore, the O-Nb-O bond angle distributions
extracted from the MD and RMC-generated configurations suggested a tendency
towards octahedral local anion co-ordination of the Nb$^{5+}$ sites \cite{Norb}. Motivated by these
considerations and by our inability to reproduce the SAED patterns in Y$_3$NbO$_7$
we have been led to examine the consequences of {\em local} cation ordering  in
this system. Note that local (rather than long-ranged) ordering presents a
significant problem for simulation studies because they necessarily involve
periodic replication of a simulation cell which may be too small to allow for the
decay of the structural correlations. Nevertheless, Miida {\it et al} \cite{Miida}
estimate the correlation length of the order responsible for the diffuse rings in
the SAED pattern to be $\sim$22 \AA\ which is the range of 4 fluorite unit cells
(the majority of our {\it d-}fluorite simulations involved 4 unit cells along each
cartesian direction, though the 11,000 atom MD cells used in the initial comparison
with SAED involved 10). Furthermore, in  the calculations to be described below,
the cation ordering is imposed upon the system, using intuition and also the
insight from the SAED studies to guess a structure which will minimise the lattice
strain. In future work we will attempt to use a Monte Carlo cation-swapping
algorithm to set up the system in a less biased way \cite{Zunger}.
\newline

The natural stoichiometry for a pyrochlore-structured oxide is $A_2$$B_2$O$_7$
where $B$ is a smaller cation which prefers octahedral coordination and $A$ a larger
one which prefers 8 oxide neighbours. We have seen above that there is a tendency
for the smaller Nb$^{5+}$ and Zr$^{4+}$ ions to preferentially bind vacancies and
hence reduce their coordination number relative to Y$^{3+}$, and adoption of the
pyrochlore arrangement would allow this tendency to be accommodated. Opposing this
tendency may be long range strain effects, such as that noted above involving
Nb$^{5+}$ in particular, and entropy, since the adoption of the pyrochlore
structure involves ordering both the cation and anion sublattices with respect to
{\it d-}fluorite. Chao Jiang {\it et al} \cite{Sick} estimate the configurational
entropy as
\begin{eqnarray}
\Delta S_{ideal}=-4k_B [x\;ln\;x+ \nonumber \\
(1-x)\;ln(1-x)+2y\;ln\;y +2(1-y)\;ln(1-y)] \label{eq1}
\end{eqnarray}
where $x$ is stoichiometric ratio of cations ($\frac {1}{2}$ for Zr$_2$Y$_2$O$_7$
and $\frac {1}{4}$ for Y$_3$NbO$_7$ ) and $y$ the fraction of vacancies to oxide
ions in the unit cell $\frac {1}{8}$. This gives configurational entropies of $4.986\times 10^{-4}$ eV/K 
and $4.534\times 10^{-4}$ eV/K  for Zr$_2$Y$_2$O$_7$ and Y$_3$NbO$_7$,
respectively. Zr$_2$Y$_2$O$_7$ has the appropriate stoichiometry to adopt the
pyrochlore structure but the radius ratio of the cations is not sufficiently large
to allow a stable, long-range ordered structure to form \cite{Sick} - certainly not
at the temperatures at which the solid-state synthesis is carried out. The
situation must be very marginal however, as Hf$_2$Y$_2$O$_7$ does form a pyrochlore
\cite{Whittle,Sick}, and the Hf$^{4+}$ ion is only slightly smaller than Zr$^{4+}$
(note that Chao Jiang {\it et al} \cite{Sick} suggest that ``covalent" effects may
play a role alongside ion size in this comparison). Some {\em local}
pyrochlore-like ordering in Zr$_2$Y$_2$O$_7$ does therefore, seem plausible. The
stoichiometry of Y$_3$NbO$_7$ is not that of a pyrochlore, but that does not rule
out a pattern of local cation and vacancy ordering which is pyrochlore-like. Note
that Irvine {\it et al} \cite{Irvine2000} consider {\em cation}-ordering in
Y$_3$NbO$_7$ likely and Miida \cite{Miida} has interpreted the SAED patterns by
proposing modulated structures in which there is some local ordering of both
vacancies and cations. The RMC-interpreted powder neutron studies \cite{Norb}
showed no evidence for cation ordering in either Zr$_2$Y$_2$O$_7$ or Y$_3$NbO$_7$,
but it must be noted that the neutron scattering lengths of Y$^{3+}$, Zr$^{4+}$, and Nb$^{5+}$ are very
similar and therefore no evidence of cation ordering can be obtained with this
technique.

\subsection{Partially-ordered Y$_3$NbO$_7$}
As mentioned above, Y$_3$NbO$_7$  does not have the correct stoichiometry to adopt a
pyrochlore structure. However, it seems reasonable to assume that the smaller
Nb$^{5+}$ ions will prefer to occupy the octahedrally-coordinated $B$ sites and we
arranged the cations so that all the Nb$^{5+}$ ions randomly occupy half of the $B$
sites with the Y$^{3+}$ occupying the remaining $B$ sites as well as all the $A$ sites.
As a consequence of the random assignment of cations to the $B$ sites the system does
not exhibit ``long"-range cation order within the simulation cell (though, because
of the simulation boundary conditions, the system is periodic). The oxide ions were
placed in the O[1] and O[2] positions.
\newline

We ran a simulation on this material in the same way as explained above on a system
with 11,000 atoms. The obtained electron diffraction spectrum is shown in figure
\ref{ED1}.  It can be appreciated that this pattern shows peaks at $G_F \pm
\frac{1}{2} (111)^*$, in reasonable agreement with the experimental data. The
reason why we do not observe two diffuse spots but a single peak is that, because
of the periodic boundary conditions, in reality we do have some long-range order in
the cation positions.
\newline

We computed the internal energies of the partially cation-ordered Y$_3$NbO$_7$ {\it
d-}fluorite at low temperatures ($T$ = 100 K). This was done by slowly cooling a
well-equilibrated high temperature run. These runs contained 704 atoms (four
fluorite unit cells in each direction). We find that the locally-ordered material
has an  energy which is lower by about $\Delta E \approx$ 56 kJ/mol = 0.58
eV/molecule \footnote{This value is found to depend weakly on the cooling rate and
on the system history.} than the {\it d-}fluorite material, confirming that some
degree of cation ordering is indeed favoured on energetic grounds. From this we can
calculate the order-disorder transition temperature, as in \cite{Sick}, $T_{O-D}
\approx \Delta E/\Delta S_{ideal} = 1160\;\rm{K}$. This value goes up to $T_{O-D}
\approx 1460\;\rm{K}$ if we assume that only 2/3 of the vacancies are ordered, {\it i.e.}
if we set $y=1/12$ in equation \ref{eq1}. Considering that this material is usually
synthesized at 1700-1800 K \cite{Norb,Miida, Irvine2000}, it seems plausible that
a certain degree of cation-ordering is observed depending upon how the sample is
cooled back to ambient conditions.
\newline

We find that the locally-ordered material is a much poorer ionic
conductor than the disordered Y$_3$NbO$_7$. We estimate \footnote{The error
associated with this number is  large because the corresponding mean squared
displacement curve is very noisy due to the low mobility of the anions.} that
cation-ordered Y$_3$NbO$_7$ has a conductivity of  $\sigma_{NE}\; \le \;
0.002\;\Omega^{-1} \rm{cm^{-1}}$ at 1500 K, approximately an order of magnitude lower
than the {\it d-}fluorite simulations at the same temperature and in much better
agreement with the experimental data. The reason for this much lower
value of the conductivity is that a partial cation ordering increases the
Nb$^{5+}$-vacancy association (see table \ref{Cat-Vac}) as well as the
vacancy-vacancy pairing along the $\langle111\rangle$ direction (see table \ref{Table III}) and
both factors hinder the vacancy mobility and therefore reduce the overall
conductivity.
 \newline

\subsection{Cation-ordering in Zr$_2$Y$_2$O$_7$}
Cation ordering in Zr$_2$Y$_2$O$_7$, on the other hand, seems  to be a less
important issue. We recall that the tendency for Zr$^{4+}$ to preferentially bind
vacancies, relative to Y$^{3+}$, is considerably lower than for Nb$^{5+}$ in Y$_3$NbO$_7$, and
the strain associated with the different local cation configurations is much less
pronounced in  this system. The experimental SAED studies show slightly different
diffuse scattering patterns \cite{Irvine2000}, \cite{Whittle}, suggesting that
lattice modulations in this material depend on the details of sample preparation.
In all cases, the diffuse features are less intense than in Y$_3$NbO$_7$. Whittle
and co-workers \cite{Whittle} have estimated a correlation length of $\approx 11$
\AA\ for the local order which is much smaller than in Y$_3$NbO$_7$. \newline

Since Zr$_2$Y$_2$O$_7$  has the correct stoichiometry for a pyrochlore, we have run
simulations starting from  a fully pyrochlore-ordered configuration ({\it i.e.} a
long-ranged ordered structure). The internal energy of this system at 100 K is
slightly lower ($\Delta E$  = 49.5 kJ/mol, a value in reasonable agreement with the
trends observed in reference \cite{Sick}) than that of a {\it d-}fluorite configuration
obtained by cooling down a high temperature run and, considering that the ordered
system has a lower entropy, it would seem that this structure is unlikely to become
stable at higher temperatures (the estimated $T_{O-D}$ is $\approx$1030 K). We also
simulated the ordered system at 1500 K to calculate the conductivity (note that
there is no cation site exchange at this temperature, so that the system cannot
relax to {\it d-}fluorite on the simulation timescale). We find the conductivity to
be very small (at least two orders of magnitude lower than the experimental value). An analysis of the
vacancy-vacancy and cation-vacancy rdfs confirms that in pyrochlore structured
Zr$_2$Y$_2$O$_7$ there is a strong pairing of the vacancies along the $\langle111
\rangle$ direction and that these bind very strongly to the Zr$^{4+}$ cations (see
table \ref{Cat-Vac} and \ref{Table III}) which explains the very small ionic
conductivity shown by this sample.

\section{Summary and Conclusions} \label{Conclusions}
We initially examined {\it d-}fluorite simulations, with a disordered cation
sublattice, which had been shown to reproduce very closely the neutron powder
diffraction data \cite{Norb}, and these already allow us to account for a
substantial part of the experimentally observed trends in the conductivity and
degree of disorder between Y$_3$NbO$_7$ and Zr$_2$Y$_2$O$_7$. The difference
between the two systems can be attributed to differences in cation size and to the
different degree of strain imposed on the lattice by the highly charged cations -
this creates oxide trapping sites in Y$_3$NbO$_7$ and gives an inhomogeneous
character to the diffusive dynamics. Although both systems exhibit a tendency to
order vacancies so that there is a likelihood of finding the vacancies close
to the smaller cation, the vacancy ordering effects are not sufficiently strong to
give rise to the diffuse scattering features which have been observed in small-area
electron diffraction studies. The fact that the observed trends are recovered from
these {\it d-}fluorite simulations implies that they are primarily associated with
the mean cation composition and traceable to the direct influence of ion size and
charge, with any influence of intermediate-range vacancy structuring an additional
feature. \newline

Because the {\it d-}fluorite simulations failed to reproduce the diffuse scattering
seen in SAED studies we were led to consider the possible effects of partial cation
ordering based upon the pattern suggested by the pyrochlore crystal structure. Our
results in this context are only indicative as we have, as yet, only examined the
properties of ``by-hand" ordered systems. Nevertheless, in Y$_3$NbO$_7$ we were
able to postulate a structure for which the diffuse scattering was in qualitative
accord with the SAED diffuse scattering patterns and which was energetically
favoured with respect to {\it d-}fluorite. In this structure the tendency of
vacancies to bind to the smaller cation was considerably enhanced relative to {\it
d-}fluorite and the conductivity was reduced by an order of magnitude, bringing it
much closer to the experimental value. For Zr$_2$Y$_2$O$_7$ we were not able to
postulate a partially ordered structure which would significantly improve upon the
{\it d-}fluorite calculations across the full range of observations (including the
SAED). In Zr$_2$Y$_2$O$_7$ the calculated conductivity in the {\it d-}fluorite
structure is already close to the experimental value and the agreement with the
powder diffraction is very good. The SAED studies indicate that the
intermediate-range order is in any case weaker in this case and our own studies on
{\it d-}fluorite show that the tendency for vacancies to associate with the smaller
cation (the driving force for pyrochlore-like ordering) is considerably reduced
relative to Y$_3$NbO$_7$. In this case a {\em fully} ordered pyrochlore is slightly
favoured energetically over the {\it d-}fluorite structure but the entropy cost is
high enough that the latter, more disordered structure, will be more stable above
$\approx$1000 K. \newline

As we have indicated above, the finding that partial cation ordering has an
influence on the material properties presents a challenge for MD simulation because
at the temperatures at which we might hope to equilibrate the system with respect
to cation positioning the ordering tendency has disappeared for entropic reasons.
However, the cation ordering also presents a challenge for comparing experimental
results. The materials are synthesised at high temperatures but then cooled
(according to different protocols) in different studies: as the temperature is
reduced the cation mobility will be drastically reduced so that the degree of
cation ordering found in low temperature studies will depart from the thermodynamic
equilibrium value to varying extents.

\section*{Acknowledgements}
We thank the EaStCHEM resource computing facility (http://www.eastchem.ac.uk/rcf)
for computing resources and the Moray Endowment Fund of the University of Edinburgh
for the purchase of a workstation. DM thanks the EPSRC, School of Chemistry,
University of Edinburgh, and the STFC CMPC for his PhD funding.  STN wishes to
thank the EU Research and Technology Development Framework Programme for financial
support.

\appendix

\section{Simulation details} \label{Sim-det}

We used dipole polarizable interaction potentials  specified in reference
\cite{Norb} (known as DIPPIM). These interaction models include full, formal
charges on the ions and the Nb$^{5+}$, Zr$^{4+}$, Y$^{3+}$, O$^{-2}$ ions have
their full, in-crystal polarizabilities. The parameters for these potentials were
obtained by ``force-matching" them to ab-initio reference data as described in
\cite{Mad1}. The full set of parameters is reported elsewhere \cite{Norb}. Most of
the simulations on Zr$_{0.5-0.5x}$Y$_{0.5+0.25x}$Nb$_{0.25x}$O$_{1.75}$ were done
using a cubic simulation box with 256 cations and 448 oxygen ions (4 x 4 x 4 unit
cells) though some calculations were done with larger cells.  The
time step used was (unless otherwise specified) 20 a.u. = $4.84\times 10^{-4}$ ps
and all the runs were performed either in an NPT or NVT ensemble, with thermostats
and barostats as described in reference \cite{Tuc1}. Most of the high temperature
runs were performed in an NVT ensemble, in which the cell volume was obtained from a
previous run in an NPT ensemble with the external pressure set to zero. Coulombic
and dispersion interactions were summed using Ewald summations while the
short-range part of the potential was truncated to half the length of the
simulation box, {\it i.e.} about 10~\AA\ . All the simulations were between 110 ps
and 10 ns long. We used an MPI version of our code which we mainly ran on a small
workstation with 2 E5462 Xeon Processors. One such processor would yield about
500,000 steps ({\it i.e.} $\approx$ 250 ps) in a day.

\section{Vacancy identification and calculation of the vacancy-cation coordination numbers} \label{Vacancy}

We can identify the positions of vacancies for an instantaneous ionic
configuration by finding which of the coordination tetrahedra are empty. Because
the cations are not diffusing we can monitor the properties of each tetrahedron
from the identities and instantaneous positions of the four cations which sit at
its vertices. Such a tetrahedron is empty if no anion is within the volume bounded
by the four planes defined by the positions of each set of three of the cations.
Because the oxide ions may perform large-amplitude vibrations about their average
sites at the temperatures of interest for dynamical studies we only assign a
vacancy to a tetrahedral site if the tetrahedron has been empty for a  minimum of
two frames (i.e.\ 10 fs). This measure has already been used for the study of PbF$_2$
\cite{Castiglione2001, Cas2} and seems to identify vacancies reasonably faithfully even at
high temperatures. For example, in simulations of Zr$_2$Y$_2$O$_7$ at 1500 K we
find that the average number of vacancies is 86 whereas the stoichiometry would
indicate 64. This number goes down to 67 at 300 K, in reasonable agreement with the
stoichiometric value. Thermally excited Frenkel pairs may make a small contribution
to the excess observed at high temperature  but the most important factor is
probably a shortcoming of our way of assigning vacancies. If a tetrahedral site
contains a vacancy, it will induce a substantial distortion of surrounding
tetrahedra which lowers the reliability of the geometric criterion we use to decide if the
site is filled or empty, especially at high temperatures where all the atoms are
making large amplitude vibrations. There is therefore a tendency to assign
vacancies to sites which neighbour a site containing a true vacancy, and this is
manifested in a larger than expected amplitude for the nearest-neighbour peak in
the vacancy-vacancy radial distribution function discussed below.
\newline

Once vacancy positions have been identified we can calculate radial distribution
functions (rdfs) between ions and vacancies and also vacancy-vacancy radial
distribution functions. Integration of the rdf can be used to define coordination
numbers. For example integrating the Y$^{3+}$-vacancy rdf, $g_{\rm{Y-V}}$ from zero out
to the position $r_c$ of first minimum of the $g_{\rm{Y-O}}$ rdf gives the average
number of vacancies in the first coordination shell of a $Y^{3+}$ ion, $n_{Y-V}$
\begin{equation}
   n_{Y-V}=4\pi \rho \int_0^{r_c} dr~ r^2g_{Y-V}(r),
 \end{equation}
where $\rho$ is the density of vacancies. Analogously, integrating the vacancy-vacancy rdf, $g_{\rm{V-V}}$ from zero out
to the position $r_c$ of first minimum of the $g_{\rm{V-V}}$ rdf gives the average
number of vacancies in the first coordination shell of another vacancy.

\section{Diffuse scattering calculation} \label{Diffuse}

 We calculate the intensity of
(total) scattering at a point ${\bf q}$ in reciprocal space from
\begin{equation}
I({\bf q})=\langle \sum_i \sum_j a_ia_j^{\star} e^{i{\bf q}\cdot{\bf r}_{ij}}
\rangle, \label{diffuse}
\end{equation}
where the sums run  over all ions in the sample, ${\bf r}_{ij}$ is the distance
between ions $i$ and $j$ and $a_i$ is the scattering amplitude of the species to
which $i$ belongs (for example, the neutron scattering length in the case of
neutron scattering). Because of the periodic boundary conditions, the accessible
vectors ${\bf q}$ with a cubic simulation cell of side $L$ are restricted to the
set ${\frac {2\pi}{L}}(m,n,p)$ where $m$, $n$, and $p$ are integers, and this
provides a limit to the resolution of the simulated pattern. \newline

We used the atomic position files obtained from the MD simulations on
{\it d-}fluorite structures to calculate the diffuse scattering from equation
\ref{diffuse}. The MD patterns were obtained from simulations on 11,000
atoms. These simulations were between 31 and 100 ps long. We started them at 2000 K
and the temperature was then lowered down to 300 K with a cooling rate lower than 
50 K/ps. The Y$^{3+}$ and Nb$^{5+}$ atoms were placed at random on the cation lattice. 
We calculated $I({\bf q})$ for the six equivalent $\langle110\rangle$ planes and averaged in 
order to show the data in the figure. We input the values of $I ({\bf q})$ on the grid 
of allowed ${\bf q}$ vectors into a contour program to generate the images shown in figure \ref{ED1}.

\end{document}